\documentclass[aps,prd,showpacs,preprintnumbers,nofootinbib,floatfix,floats,groupedaddress,twocolumn]{revtex4}
\usepackage{bm}
\usepackage{latexsym}
\usepackage{dcolumn}
\usepackage{amsmath,amsfonts,amssymb}
\usepackage{graphicx,epsfig}

\def\eq#1{{Eq.~(\ref{#1})}}

\reversemarginpar

\begin{document}
\title{Why Does the Universe Expand ?}
 \author{T.~Padmanabhan}
 \email{paddy@iucaa.ernet.in}
 \affiliation{IUCAA,
 Post Bag 4, Ganeshkhind, Pune - 411 007, India}

\date{}
\begin{abstract}
The purpose of the paper is five-fold: (a) Argue that the question in the title can be presented in a meaningful manner and that it requires an answer. (b) Discuss the conventional answers and explain why they are unsatisfactory. (c) Suggest that a key ingredient in the answer could be the instability arising due to the `wrong' sign in the Hilbert action for the kinetic energy term corresponding to expansion factor. (d) Describe how this idea connects up with another peculiar feature of our universe, viz. it spontaneously became more and more classical  in the course of evolution. (e) Provide a speculative but plausible scenario, based on the thermodynamic perspective of gravity, in which one has the hope for relating the thermodynamic and cosmological arrows of time.

\end{abstract}

\maketitle
\vskip 0.5 in
\noindent
\maketitle

\def\eq#1{{Eq.~(\ref{#1})}}
\def\eqs#1{{Eqs.~(\ref{#1})}}
\def\fig#1{{Fig.~\ref{#1}}}
\def\figs#1{{Figs.~\ref{#1}}}


Observations show  that our universe, at sufficiently large scales, is homogeneous and isotropic and is described by a Friedmann model. The dynamics of such a model is contained in a single function of time $a(t)$, called expansion factor.\footnote{To be precise we also have to specify the curvature of spatial slices by $k=0,\pm 1$ but we will set $k=0$ for simplicity. None of the key results in the paper depend on this choice, which, in any case, is favored observationally.}
Other degrees of freedom in the metric tensor, needed to describe the universe at smaller scales, do not play any role at sufficiently large scales. Therefore, one can model the universe as a dynamical system governed by differential equations determining the evolution of $a(t)$. These, obtained from Einstein's equations, are:
\begin{equation}
\left(\frac{\dot a}{a}\right)^2=\frac{8\pi L_P^2}{3}\rho(t)
\label{evone}
\end{equation}
and
\begin{equation}
\frac{\ddot a}{a}=-\frac{4\pi L_P^2}{3}(\rho+3p)
\label{evtwo}
\end{equation} 
where $\rho$ and $p$ are the energy density and pressure of matter. (The units are chosen with $c=\hbar=1$ so that $L_P^2=G$). We will assume, for simplicity, that there is only one component of matter with some equation of state $p=p(\rho)$. These two equations imply (when $\dot a\neq 0$) that $d(\rho a^3)=-p d(a^3)$.
To determine the dynamics from \eq{evtwo}, we need to specify $(\dot a, a,\rho)$ at some fiducial time $t=t_{fid}$ such that they are consistent with the constraint \eq{evone}. This will allow us to solve the equations and determine the dynamics of the universe for both $t<t_{fid}$ and $t>t_{fid}$.

Equation (\ref{evone}) and \eq{evtwo} are invariant under the time reversal $t\to -t$. 
It turns out that, to match with observations, we have to choose $\dot a>0$ at $t=t_{fid}>0$ (say, at the current epoch), thereby breaking the time-reversal invariance of the system. This, by itself, is not an issue for a physical system in the lab. We know that specific solutions to dynamical equations need not respect the symmetries of the
equations. But, for the universe, this \textit{is} an issue. 

To see why, let us first consider the case of $(\rho+3p)>0$ for all $t$.
The choice $\dot a>0$ implies that we are \textit{assuming} that the universe is expanding. Then \eq{evone}, \eq{evtwo} show that the universe will expand at all times in the past and will have a  singularity ($a=0$) at some finite time in the past (which we can take to be $t=0$ without loss of generality). The mathematical structure of \eq{evone} and \eq{evtwo} prevents us from specifying the initial conditions at $t=0$. We are therefore forced to take $\dot a>0$ at some $t=\epsilon>0$, thereby breaking time reversal symmetry. The universe expands at present because we choose it to expand at some epoch in the past for reasons we do not know. 
This expansion, in turn, leads to an arrow of time
such that either $a$ or $t$ can be used as a time coordinate --- which is precisely what a cosmologist does in using redshift $z=(1/a)-1$ to specify the cosmic epoch.
Why do we have to choose the solution with $\dot a>0$ at some epoch?. This is the essence of \textit{expansion problem} \cite{peacock}.
An alternative way of asking the same question is: How come a cosmological arrow of time\footnote{The phrase `cosmological arrow of time' means different things to different people; I will use it in the sense that $a(t)$ being a monotonic function of $t$ with $\dot a>0$,  gives a direction for $t$ from the  evolution of $a$.} emerges from equations of motion which are time-reversal invariant ?

In a laboratory system, we can usually take another copy of the system and explore it with
a time-reversed choice of initial conditions, because the time can be specified by degrees of freedom external to the  system. We cannot do it for the universe because we do not have extra copies of it and --- equally importantly --- there is nothing external to it to specify the time. So the problem, as stated, is specific to cosmology.

In the above discussion we assumed $(\rho+3p)>0$ thereby leading to a singularity. Since physically meaningful theories must be nonsingular, we certainly expect a future theory of gravity --- possibly a model for quantum gravity --- to eliminate the singularity (effectively leading to ($(\rho + 3p) <0$). Could it be that such a theory might solve the problem of arrow of time? This is unlikely.  To see this, let us ask what kind of dynamics we expect in such a `final' theory. The dynamics will certainly get modified at Planck epoch but away from it we expect some effective equations (possibly with quantum corrections) to govern the evolution of an (effective) expansion factor. The solutions could, for example, have a contracting phase (and bounce) or could start from a Planck size universe at $t=-\infty$, just to give two non-singular possibilities. While we do not know these equations or solutions, we can be confident that they will still be time-reversal invariant because quantum theory, as we know it, is time-reversal invariant. So without a choice for initial conditions (now possibly at $t=-\infty$), we still cannot explain how the cosmological arrow of time emerges. Since quantum gravity is unlikely produce an arrow of time, it is worthwhile to attempt to understand this problem in the (semi)classical context.

There is another peculiarity related to the quatum regime of the universe which does not seem to have attracted the attention it deserves. The conventional wisdom is that the universe was a quantum system with order unity quantum fluctuations near Planck time. But during the evolution it seems to have become more and more classical! \textit{This is an extraordinary feature} and again needs explanation. In general,  systems with bounded Hamiltonians do not become more and more classical\footnote{It is possible to quantify the degree of classicality of the system using, e.g., the Wigner function and specifying how sharp is the phase space trajectory.} \textit{spontaneously} under dynamical evolution. One suspects that this peculiarity is related to the origin of cosmic arrow of time. 

A more complicated ``solution" to arrow of time issue, which is sometimes envisaged, is as follows: 
Consider a very inhomogeneous initial condition in some $t=t_i$ hypersurface and assume that
certain regions behave like `local' Friedmann models with $\dot a>0$ and other regions have
$\dot a<0$ so that no global arrow of time can be defined from the expansion. Further assume that the dynamics of these patches are independent of each other and we just happen to be in a patch with 
$\dot a>0$ thereby `solving' the problem. There are several difficulties with this scenario. It is impossible to prove that, for a generic initial condition, the patches will evolve independently in a nonlinear theory of gravity and --- in fact --- it is very unlikely to be true. Even defining a sensible `local expansion factor' for a `local patch' without assuming special symmetries is impossible. Such scenarios are often invoked in the context of inflationary models but not with mathematically rigorous proofs.

It will be nicer and simpler if we can find a way by which the  equations of motion that are time-reversal invariant  lead to  an evolution which singles out an arrow of time. At first sight one might think  this is impossible but one can do it with unbounded Hamiltonians. Let me first describe the idea with a simple example. Consider an `inverted' harmonic oscillator $q(t)$ obeying the equations of motion
\begin{equation}
\ddot q=\omega^2 q
\label{invosci}
\end{equation} 
This equation is invariant under $t\to -t$ so one would have thought that no arrow of time will emerge from the dynamics unless we impose it in the initial conditions. The general solution to this equation is
\begin{equation}
q(t)=q(0)\cosh\omega t+\dot q(0)\omega^{-1}\sinh\omega t
\label{mirone}
\end{equation}
For generic initial conditions, there is no relationship between $q(0)$ and $\dot q(0)$. Hence at late times (i.e,
$t\gg \omega^{-1}$), we get one part of the solution being selected out
\begin{equation}
q(t)\approx\frac{1}{2}(q(0)+\dot q(0)\omega^{-1})e^{\omega t}\propto e^{\omega t}
\end{equation} 
leading to an ``expansion" and an arrow of time! Note that the solution in \eq{mirone} 
is time-reversal invariant in the sense that $q(t)=q(-t)$ if we let $\dot q(0)\to -\dot q(0)$ when we do $t\to-t$. But once we have chosen any generic solution with some uncorrelated $q(0)$ and $\dot q(0)$, late time dynamics picks out an arrow of time. (Of course, there are special initial conditions like $q(0)=-\dot q(0)\omega^{-1}$ or $q(0)=0=\dot q(0)$ for which this will not happen but these are special choices and not generic.)

One can easily see that 
 this behaviour arises for a wide class of 
 Hamiltonians that are unbounded. It is not necessary that the potential energy has to be unbounded. If the kinetic energy term has the `wrong' sign so that the Lagrangian has a form like $L=-(1/2)\dot q^2-V(q)$ with a $V$ which is positive and unbounded from above, say, we will again have an instability and late time evolution of $q$ will give an arrow of time.

Incredibly enough, it is known for decades \cite{bryce} that the expansion factor $a(t)$ does have such a wrong sign in the kinetic energy term in the Hilbert action and hence represents an unstable mode. Our analysis suggests that it is this cosmic instability which we call expansion, and for timescales larger than Planck time, it picks out an arrow of time. Clearly, same feature will occur even in any effective theory of (semi)classical gravity once $a(t)$ acquires an unstable dynamics. This is guaranteed to happen because any sensible theory will approach Friedmann model at times larger than Planck time and --- in the Friedmann limit --- $a(t)$ is an unstable model.

There are two other nice features in this attempt to connect the cosmic arrow of time with the wrong-sign kinetic energy term for $a(t)$ and the consequent instability. 

First, the idea tells us that it is very special to cosmology. Unbounded Hamiltonians with negative kinetic energy term are taboo in laboratory scale physics for good reasons; one does not want run-away solutions in the lab systems. But one can easily accommodate such an instability in the behavior of the universe. The cosmic expansion is just a run-away solution fed by the instability.

Second, this idea ties up (i)  the origin of cosmic arrow of time to (ii) the quantum to classical transition made by the universe during its evolution. As we said before, normal lab systems do not spontaneously evolve into a more and more classical state (in the sense of, say, the Wigner function becoming more and more sharply peaked around the classical trajectory) as time evolves. But this is precisely what systems with unbounded Hamiltonians do. For example, the inverted harmonic oscillator in \eq{invosci}, treated as a quatum system, is known to become more and more classical as it evolves \cite{guthandpi}.

More generally, suppose we decompose the spatial 3-metric in the form $g_{\alpha\beta}=a^2h_{\alpha\beta}$ with $\mathrm{det}\ h=1$ as a gauge condition. Using the form of Einstein-Hilbert Lagrangian, one can show that while the kinetic energy term for $a(t)$ has the wrong-sign, the other degrees of freedom, represented by $h_{\alpha\beta}$, have the correct sign. In other words, it is only the overall scale factor which has an instability. Hence it is this mode which turns classical first during the evolution. If these features are maintained in the effective quantum-corrected description of gravity, then we can hope to have an explanation for a broader question: Why is the classical universe described by a single dynamical degree of freedom, rather than by, say a Bianchi type-I model with three degrees of freedom?

In addition to the cosmic arrow of time determined by the dynamics of the expansion factor, we also have another well-known arrow of time, determined by the direction of increase of entropy. The question arises as to whether the cosmological dynamics can be given a thermodynamic interpretation so that the expansion of the universe can be linked to the growth of gravitational entropy. Obviously the answer depends on the definition used for the gravitational entropy and one idea --- originally due to Penrose \cite{penrose} --- is to relate the gravitational entropy to Weyl curvature, $C^{ab}_{ij}$. Since $C^{ab}_{ij}=0$
for the Friedmann universe and becomes non-zero as inhomogeneities develop, this approach links the  cosmic arrow of time with the direction of increasing gravitational entropy.

This approach, however, requires us to move outside the realm of Friedmann model; that is, a strictly Friedmann model will remain a zero entropy (or constant entropy) system in this framework even though the expansion does give it a cosmological arrow of time. From this perspective, it may be nicer if we can define the gravitational entropy in some other --- meaningful --- way so that one can work within the framework of Friedmann models without having to introduce the inhomogeneities.

That such an interpretation should be possible is indicated by the fact that the Einstein's equations themselves can be obtained from an entropy maximization principle \cite{aseementropy}.
There has also been extensive work relating the Friedmann equations to the thermodynamic identity $TdS=dE+PdV$ and its variants  \cite{ronggen}. 
All these suggest that gravity is an emergent phenomenon and one should be able to associate gravitational entropy with the (unknown) microscopic degrees of freedom underlying the spacetime geometry.

Somewhat ironically, while it is possible to obtain the \textit{full set} of Einstein's equations from an entropy maximization procedure \cite{aseementropy}, it is not easy to apply this technique to a restricted class of metrics (like the maximally symmetric spacetimes) and obtain the relevant equations of motion in that context. Such a restriction works naturally only in the case of static spacetimes. So this idea works very well, for example,  for de Sitter
universe (which can be expressed in a static coordinate system) but not in general. 

I will now discuss to what extent one may be able to use the thermodynamic perspective of gravity and link the cosmic arrow of time to the direction of increasing gravitational entropy. 
In ref. \cite{cqgpap}, I introduced a definition for entropy $S$ associated with a two-surface, $\partial \mathcal{V}$, bounding a three-volume
$\mathcal{V}$, in static spacetimes with a horizon,  by (see eq. (2) of ref. \cite{cqgpap}):
\begin{equation}
TS 
= \int_{\partial\cal V}\frac{\sqrt{\sigma}d^2x}{4L_P^2}\left(
\frac{N|n_\mu a^\mu|}{2\pi}\right)\equiv\frac{1}{4}\int_{\partial\cal V} dn T_{loc}
\label{defs}
\end{equation}
Here $a^i=u^j\nabla_ju^i=(0,\partial^\mu N/N)$ is the acceleration of
the $x^\mu=$ constant observers with four-velocity $u^i$, $N$ is the lapse function with $N=0$ representing the horizon $\mathcal{H}$ with temperature $T=|\kappa|/2\pi$ where $|\kappa|$ is the magnitude of the surface gravity.
(The issue of signs in these expressions is discussed in ref.\cite{cqgpap}.) 
 The second equality in \eq{defs} identifies the number of (microscopic) degrees of freedom in an area element $\sqrt{\sigma} \, d^2 x$ to be $dn=  \sqrt{\sigma}\, d^2x/L_P^2$ and the local Unruh temperature related to the acceleration to be $T_{\rm loc}\equiv (N |a^\mu n_\mu|) /2\pi$.
Using the expression for active gravitational mass-energy  defined by (see eq. (8) of ref. \cite{cqgpap}) the Komar-Tolman integral:
\begin{equation}
\label{defe}
E= 2\int_{\cal V}d^3x\sqrt{\gamma}N (\bar T_{ab}u^au^b)
\end{equation}
where $\bar T_{ab}\equiv(T_{ab}-\frac{1}{2}Tg_{ab})$ and using Einstein's equations one can immediately obtain the relation
\begin{equation}
S=\frac{E}{2T} \equiv \frac{1}{2} \beta E
\label{one}
\end{equation} 
This was obtained and discussed in detail in Ref. \cite{cqgpap}. When combined with \eq{defs},
this result can be expressed  as the `equipartition law' $\Delta E= (1/2) (k_B T_{\rm loc}) \Delta n$ for microscopic degrees of freedom located in the area elements of  ${\partial\cal V}$. Alternatively, one can obtain Einstein's equations from \eq{one} by reversing the steps. (In fact, all these work for a wider class of theories called Lovelock gravity, of which Einstein's theory is a special case; see \cite{equiparti}). For example, in the case of deSitter universe (which has a well-defined thermodynamics \cite{GH} and a static coordinate representation with a horizon)
discussed in ref.\cite{cqgpap}, this procedure is straightforward.

For a general Friedmann model, however, we will not have a coordinate system in which one can define horizon, effective Unruh temperature etc. A possible way of using \eq{one} in this context is suggested by the fact its derivation in ref.\cite{cqgpap} used the Einstein's equations in the form $\nabla_\mu a^\mu\propto \bar T_{ab}u^au^b$ where $a^i$ was the proper acceleration of observers with $x^\mu$= constant. This  will not hold in the case of a general time-dependent metric; in fact, for Friendman universe the proper acceleration of comoving observers is zero. But we know that the relation $\nabla_\mu a^\mu\propto \bar T_{ab}u^au^b$ will continue to hold in \textit{any} spacetime, if we take $a^\mu$ to be the proper acceleration of the \textit{geodesic deviation vector} and the expressions are evaluated in the comoving frame. This suggests that one could try to interpret \eq{one} for Friedmann universe with $a^\mu$ identified with the acceleration of the geodesic separation vector. 

Choosing  ${\partial\cal V}$  to be a two-sphere of fixed proper radius $R=r a(t)$, it seems reasonable to set $E=(\rho+3p)(4\pi R^3/3)$ and  $S=A/4L_P^2$ where $A\propto a^2$  is the proper area of ${\partial\cal V}$. (This gives  $\dot S/S=2\dot a/a$ for each surface and is consistent with the expansion of the universe leading to increase in  gravitational entropy, but in a rather naive way.) If we further identify $T$  with the Unruh temperature corresponding to the \textit{magnitude} of the  acceleration $|(\ddot a/a)|R$ of the geodesic deviation vector, then \eq{one} will reduce to \eq{evtwo} as far as magnitude of the acceleration is concerned. (The $r$ drops out of the relation and hence ${\partial\cal V}$ could have been chosen to have special properties, like e.g, location of horizon etc. if desired.) One cannot determine  the (crucial) fact that the universe is decelerating for $(\rho+3p)>0$ without further assumptions since Unruh temperature only depends on the magnitude of the acceleration. Unfortunately, there is no simple justification for defining $T$ using the acceleration of geodesic deviation vector and hence the success of the procedure must be considered fortuitous at this stage.

More rigorously, one can express any Friedmann model in a spherically symmetric form with a time-dependent metric. Making several further choices (like e.g., Kodama vector in the absence of Killing vector to define energy) one can relate Einstein's equations with horizon thermodynamics, as has been done in the literature \cite{ronggen} in many ways. Reversing these steps, one can obtain Einstein's equations from the thermodynamics of horizons, provided one makes specific choices. 

While the details of such choices differ significantly, this programme provides the roadmap for linking the cosmic arrow of time with the thermodynamic arrow of time in  four steps: First, we connect the cosmic arrow of time to the dynamical evolution governed by Einstein's equations, which was the main scope of this paper. Second, we link the gravitational field equations to the thermodynamics of horizons for which we have fair amount of results in general \cite{aseementropy,equiparti} as well as in the context of cosmology \cite{ronggen}.
 But in the latter context one needs to understand better the physical origin of various choices, in a time dependent scenario. Third, we link horizon thermodynamics to `normal' matter entropy using the physical processes version of first law of horizon and a suitable form of generalized second law. This is again nontrivial. The rigorous results are only available for blackhole spacetimes with asymptotic flatness, stationarity etc. while we need them in the case of time-dependent, observer-dependent, cosmological horizons. One requires a local formulation  which does not exist at present.
Finally we link the normal entropy and second law of thermodynamics to the thermodynamic arrow of time. Successful completion of these steps will provide a link between the two arrows of time.

\textit{Acknowledgements}: These ideas were first presented in my lecture at ICTS-IUCAA Programme on Cosmology with CMB and LSS, July-Aug 2008 and I thank the participants for their feedback. I also thank H.Nicolai and R.Penrose for useful discussions on this topic during the `Foundations of Space and Time' meeting at Cape Town, 10-14 August 2009.


 \end{document}